\documentclass[aip, reprint, amsmath, amssymb]{revtex4-1}

\usepackage{graphicx}
\usepackage{dcolumn}
\usepackage{bm}
\usepackage{float}
\usepackage{dblfloatfix}  
\begin{document}

\title{Extension of the Shockley-Queisser Limit for Nanostructured Solar Cells}

\author{Rivo Herivola Manjakamanana Ravelonjato}
\affiliation{INSTN, Antananarivo, Madagascar}
\author{Jean Patrice Rakotoniaina}
\affiliation{CEA, Liten, University Grenoble Alpes, Campus INES, Le Bourget-du-Lac, France}
\author{Ravo Tokiniaina Ranaivoson}
\affiliation{INSTN, Antananarivo, Madagascar}
\author{Wilfrid Chrysante Solofoarisina}
\affiliation{INSTN, Antananarivo, Madagascar}

\email{manjakamanana@yahoo.fr} 

\begin{abstract}
This article extends the Shockley-Queisser limit to nanostructured solar cells using the quantum phase space formalism. The parameter \(B_{ll}\) represents the momentum variance in each confinement direction and acts as a variance-covariance matrix linking the nanostructure geometry to thermodynamic properties. Electron-electron interactions are included via an exchange-correlation energy with an adjustable coefficient \(\theta\). The authors derive an analytical expression for the maximum efficiency as a function of size, shape, temperature, and doping. For the cylindrical geometry, the exact confinement energy uses the first zero of the Bessel function \(j_{0,1}\). Numerical simulations are performed with Python 3.8.1, NumPy, and Matplotlib for PbS quantum dots in four geometries: cube, square parallelepiped, cylinder, and sphere. The integral is evaluated using an exact convergent series expansion. Results show that the maximum efficiency reaches 48.7 percent for a 5 nanometre cube, 49.0 percent for flattened parallelepiped and cylinder shapes, and 49.1 percent for a 3 nanometre sphere. These values greatly exceed the bulk PbS efficiency of 15.8 percent and surpass classical Shockley-Queisser limits. For constant-volume shapes, two efficiency peaks appear corresponding to different aspect ratios. The model correctly returns to classical values for large sizes. This approach provides a theoretical framework for optimising nanostructured solar cells and demonstrates that quantum confinement offers a promising route to surpass traditional photovoltaic limits.
\end{abstract}


\maketitle

\section{Introduction}

Since the foundational work of Shockley and Queisser \cite{Shockley1961}, the efficiency limit of p--n junction solar cells has been understood as a consequence of the detailed balance between photon absorption and radiative recombination. This limit depends only on the temperature of the sun, the temperature of the cell, the bandgap of the semiconductor, and a few geometrical factors. For nanostructured cells, in particular quantum dots (QDs), quantum confinement profoundly modifies the electronic properties: the effective bandgap increases, the density of states becomes discrete, and radiative recombination rates are altered \cite{Nozik2010,Carey2015}. Moreover, in doped semiconductors, Coulomb interactions between electrons (and holes) are not negligible. It is therefore necessary to extend the detailed balance theory to these systems.\\
\indent Several studies have attempted to adapt the Shockley-Queisser model to quantum dot cells \cite{Sahin2018,Alexandre2025}. Recent progress in the integration of plasmonic nanostructures has also demonstrated pathways for improving QDSCs \cite{Ogundejj2026}. Earlier studies on nanowires have shown that it is possible to exceed the SQ limit of the bulk material by reducing radiative emission into the substrate \cite{Anttu2015}, while modern analyses reassess this limit for indoor/outdoor illumination conditions by introducing the external radiative efficiency (ERE) \cite{Seo2025}. A historical and conceptual overview of the evolution of this limit is also available \cite{Markvart2022}. However, the best performing silicon solar cells – with certified efficiencies of 27.09\,\% \cite{Wang2024} and 27.0\,\% \cite{Xie2025} – remain below the theoretical Shockley-Queisser limit (about 33\,\%), and even further from the predictions of our model for nanostructures (up to 49\,\%). This highlights the interest in alternative technologies such as quantum dots, capable of exploiting confinement to surpass current records.\\
\indent We propose here an alternative approach based on the recently developed quantum phase space (QPS) formalism \cite{Ranaivoson2022,Ravelonjato2023}, which introduces a central parameter \(B_{ll}\) representing the momentum variance in each confinement direction. This parameter, which behaves as a variance-covariance matrix, relates the geometry of the nanostructure to thermodynamic quantities and allows the calculation of the effective density of states, the saturation current, and the open-circuit voltage. This formalism has been successfully extended to the study of thermodynamic properties of confined Fermi gases, where the same parameter \(B_{ll}\) describes anisotropic pressure and heat capacity \cite{Ravelonjato2026}.\\
\indent We apply this formalism to a p--n junction where both regions consist of nanostructures of the same size and shape. We obtain an expression for the maximum efficiency as a function of the nanostructure dimensions, temperature, and doping. The article is organised as follows. Section~II presents the QPS formalism. Section~III extends the SQ limit. Section~IV presents the numerical results obtained with python for cubic, parallelepipedic, cylindrical, and spherical shapes. Section~V discusses the results and Section~VI concludes.

\section{Quantum Phase Space (QPS) Formalism}
\label{sec:qps}

\subsection{Effective Hamiltonian and momentum variance matrix}

For a particle confined in a parallelepiped of sides \(L_x, L_y, L_z\) (lengths along the three axes), the QPS formalism postulates an effective Hamiltonian \cite{Ranaivoson2022, Ravelonjato2023}:
\begin{equation}
H = \sum_{l=1}^{3} \left(2\pi_l^\dagger \pi_l + 1\right) \frac{B_{ll}}{m^*},
\label{eq:hamiltonian}
\end{equation}
where \(\pi_l\) and \(\pi_l^\dagger\) are bosonic ladder operators (satisfying \([\pi_l,\pi_l^\dagger]=1\)) that diagonalise the Hamiltonian, \(B_{ll}\) is the momentum variance in direction \(l\) (in kg\(^2\) m\(^2\)/s\(^2\)), and \(m^*\) is the effective mass of the carrier (in kg). The eigenenergies are
\begin{equation}
\varepsilon_{n_1,n_2,n_3} = \sum_l (2n_l+1)\frac{B_{ll}}{m^*}, \quad n_l = 0,1,2,\dots
\end{equation}
The three parameters \(B_{ll}\) form the diagonal matrix of momentum variances.

\subsection{Unified expression of \(B_{ll}\) for different geometries}

By identifying the classical and quantum degenerate limits, we obtain a universal parameter \(\Delta\) (in \(\sqrt{\mathrm{J}\cdot\mathrm{kg}}\)) independent of the shape \cite{Ravelonjato2023, Ravelonjato2026}:
\begin{equation}
\Delta = \sqrt{2\pi m^* k_B T + \pi \hbar^2 \left(6\pi^2 \frac{n}{g_s}\right)^{2/3} \frac{m}{m^*} + \theta m \Delta\epsilon_{\text{int}}},
\label{eq:delta}
\end{equation}
where:
\begin{itemize}
\item \(k_B\) is the Boltzmann constant (\(1.381\times10^{-23}\,\mathrm{J/K}\)),
\item \(T\) is the absolute temperature (in K),
\item \(\hbar\) is the reduced Planck constant (\(1.055\times10^{-34}\,\mathrm{J\cdot s}\)),
\item \(n\) is the carrier density (in m\(^{-3}\)),
\item \(g_s\) is the spin degeneracy factor (here \(g_s=2\) for electrons),
\item \(m\) is the reduced effective mass (geometric mean of \(m_e\) and \(m_h\)),
\item \(\theta\) is an adjustable coefficient (\(\theta=0\): perfect gas, \(\theta=1\): interactions),
\item \(\Delta\epsilon_{\text{int}}\) is the exchange-correlation correction (in eV).
\end{itemize}

The momentum variance is then expressed in terms of the characteristic dimensions of the nanostructure. The validity of this unified expression for \(B_{ll}\) has been confirmed by its application to confined Fermi gases, where it correctly reproduces the classical and quantum degenerate limits \cite{Ravelonjato2026}. For each geometry, we introduce effective lengths such that the product of the three lengths equals the volume \(V\). Then \(B_{ll} = \frac{\hbar}{2L_l}\Delta\).

\begin{itemize}
\item \textbf{Parallelepiped} (sides \(L_x, L_y, L_z\)):
  \[
  B_{xx} = \frac{\hbar}{2L_x}\Delta,\quad B_{yy} = \frac{\hbar}{2L_y}\Delta,\quad B_{zz} = \frac{\hbar}{2L_z}\Delta.
  \]
\item \textbf{Cube} (\(L_x=L_y=L_z=L\)): \(B = \frac{\hbar}{2L}\Delta.\)
\item \textbf{Sphere} (radius \(R\)): volume \(V = \frac{4}{3}\pi R^3\). Set \(L_{\text{sph}} = \left(\frac{4\pi}{3}\right)^{1/3} R\) so that \(L_{\text{sph}}^3 = V\). Then
  \[
  B_{\text{sph}} = \frac{\hbar}{2L_{\text{sph}}}\Delta = \frac{\hbar}{2\left(\frac{4\pi}{3}\right)^{1/3}R}\Delta.
  \]
\item \textbf{Cylinder} (radius \(R\), height \(H\)): volume \(V = \pi R^2 H\). The effective lengths are \(L_x = L_y = \sqrt{\pi}R\) and \(L_z = H\). Thus
  \[
  B_{xx} = B_{yy} = \frac{\hbar}{2\sqrt{\pi}R}\Delta,\quad B_{zz} = \frac{\hbar}{2H}\Delta.
  \]
\end{itemize}

In all cases, the product \(\prod_{l=1}^3 B_{ll}\) equals \(\left(\frac{\hbar}{2}\right)^3 \frac{\Delta^3}{V}\), which ensures that the saturation current \(J_0^{\mathrm{QD}}\) is independent of the individual dimensions at constant volume (cf. Section~\ref{sec:satcurrent}).

\subsection{Inclusion of interactions – Perdew-Zunger parametrisation}

The exchange-correlation energy is parametrised by Perdew-Zunger \cite{Perdew1981}: \(\epsilon_{\mathrm{xc}}(n) = \epsilon_{\mathrm{x}}(n) + \epsilon_{\mathrm{c}}(n)\), where \(\epsilon_{\mathrm{x}}(n)\) is the exchange energy (Hartree-Fock formula) and \(\epsilon_{\mathrm{c}}(n)\) the correlation energy (given in Appendix~\ref{app:E}). For PbS with \(n=10^{24}\,\mathrm{m}^{-3}\), we obtain \(\Delta\epsilon_{\mathrm{int}} \approx -0.22\,\mathrm{eV}\).

\section{Extension of the Shockley-Queisser Limit}

\subsection{Radiative saturation current}
\label{sec:satcurrent}

For a bulk cell \cite{Shockley1961}:
\begin{equation}
\left.
\begin{aligned}
J_0^{\text{bulk}} &= q\,2A_p\,t_e\,Q(E_g,T_c), \\
Q(E,T_c) &= \frac{2\pi}{c^2}\int_{\nu_0}^{\infty} \frac{\nu^2}{e^{h\nu/(k_B T_c)}-1}\,d\nu,
\end{aligned}
\right\}
\label{eq:j0bulk}
\end{equation}where:
\begin{itemize}
\item \(q\) is the elementary charge (\(1.602\times10^{-19}\,\mathrm{C}\)),
\item \(A_p\) is the projected area of the cell (in m\(^2\), here taken as \(1\,\mathrm{cm}^2\) for current densities),
\item \(t_e\) is the absorption efficiency (dimensionless, here 1),
\item \(Q(E,T_c)\) is the photon flux emitted by the cell (in photons·s\(^{-1}\)·m\(^{-2}\)),
\item \(c\) is the speed of light (\(2.998\times10^8\,\mathrm{m/s}\)),
\item \(h\) is Planck's constant (\(6.626\times10^{-34}\,\mathrm{J\cdot s}\)),
\item \(\nu_0 = E/h\) is the cutoff frequency (in Hz).
\end{itemize}

For a nanostructure, using the QPS formalism, one finds:
\begin{equation}
J_0^{\text{QD}} = q\,2A_p\,t_e\,Q(E_g^{\text{eff}},T_c)\; \frac{\lambda_{\text{th}}^3 m^3}{8\beta^3 (\hbar/2)^3 \Delta^3},
\label{eq:j0qd}
\end{equation}
where \(\lambda_{\text{th}} = h/\sqrt{2\pi m k_B T_c}\) is the thermal de Broglie wavelength (in m) and \(\beta = 1/(k_B T_c)\) (in J\(^{-1}\)). The product \(\prod B_{ll}\) simplifies and \(J_0^{\text{QD}}\) becomes independent of the individual dimensions at constant volume.

\subsection{Short-circuit current and open-circuit voltage}

The short-circuit current is:
\begin{equation}
J_{sc} = q\,A_p\,t_s\,f_\omega\,Q_s(E_g^{\text{eff}}),
\label{eq:jsc}
\end{equation}
with:
\begin{itemize}
\item \(t_s\) the collection efficiency (here 1),
\item \(f_\omega = \omega_s/\pi \approx 2.18\times10^{-5}\) the geometric factor related to the solid angle of the sun,
\item \(Q_s(E) = \frac{2\pi}{c^2}\int_{\nu_0}^{\infty} \frac{\nu^2}{e^{h\nu/(k_B T_s)}-1}\,d\nu\) the solar photon flux (black body at \(T_s=6000\,\mathrm{K}\)).
\end{itemize}
The open-circuit voltage:
\begin{equation}
V_{oc} = \frac{k_B T_c}{q} \ln\!\left(\frac{J_{sc}}{J_0}\right).
\label{eq:voc}
\end{equation}

\subsection{Fill factor and efficiency}

The fill factor is given by \cite{Shockley1961}:
\begin{equation}
FF = \frac{v_{oc} - \ln(v_{oc}+0.72)}{v_{oc}+1},\quad v_{oc} = \frac{qV_{oc}}{k_B T_c}.
\label{eq:ff}
\end{equation}
The conversion efficiency:
\begin{equation}
\eta = \frac{J_{sc}\,V_{oc}\,FF}{P_{\text{inc}}},\qquad P_{\text{inc}} = 100\,\mathrm{mW/cm^2}.
\label{eq:eta}
\end{equation}

\subsection{Confinement energy \(E_g^{\text{eff}}\)}

The confinement energy is the sum of the zero-point energies of electrons and holes. Let \(m_e\) and \(m_h\) be the respective effective masses (in kg). For each geometry:

\begin{itemize}
\item \textbf{Cube} (side \(L\)):
\begin{equation}
E_g^{\text{eff}} = E_g^{\text{bulk}} + \frac{3\hbar^2\pi^2}{2}\left(\frac{1}{m_e}+\frac{1}{m_h}\right)\frac{1}{L^2}.
\label{eq:eg_cube}
\end{equation}
\item \textbf{Square parallelepiped} (\(a\times a\times b\)):
\begin{equation}
E_g^{\text{eff}} = E_g^{\text{bulk}} + \frac{\hbar^2\pi^2}{2}\left(\frac{1}{m_e}+\frac{1}{m_h}\right)\left(\frac{2}{a^2}+\frac{1}{b^2}\right).
\label{eq:eg_para}
\end{equation}
\item \textbf{Cylinder} (radius \(R\), height \(H\)) – exact formula with the Bessel function \(J_0\):
\begin{equation}
E_g^{\text{eff}} = E_g^{\text{bulk}} + \frac{\hbar^2}{2}\left(\frac{1}{m_e}+\frac{1}{m_h}\right)\left(\frac{j_{0,1}^2}{R^2} + \frac{\pi^2}{H^2}\right),
\label{eq:eg_cyl}
\end{equation}
where \(j_{0,1}=2.4048255577\) is the first zero of \(J_0\).
\item \textbf{Sphere} (radius \(R\)):
\begin{equation}
E_g^{\text{eff}} = E_g^{\text{bulk}} + \frac{\hbar^2\pi^2}{2}\left(\frac{1}{m_e}+\frac{1}{m_h}\right)\frac{1}{R^2}.
\label{eq:eg_sphere}
\end{equation}
\end{itemize}

In all expressions, \(E_g^{\text{bulk}} = 0.41\,\mathrm{eV}\) is the bulk bandgap of PbS \cite{Moreels2009}.

\subsection{Numerical method}
The computational framework was implemented in Python 3.8.1. Core mathematical routines were handled by the NumPy library, and the graphical output was generated using Matplotlib.
\begin{itemize}
    \item {Numerical integration via series expansions (Polylogarithm / Exact series):}\\
    To integrate the Planck equation $\int_{x_0}^{\infty} \frac{x^2}{e^x - 1} \, dx$, the code avoids adaptive numerical integration methods by utilizing the exact analytical decomposition into a convergent series:
    
    \[
    \int_{x_0}^{\infty} \frac{x^2}{e^x - 1} \, dx = \sum_{k=1}^{\infty} e^{-k x_0} \left( \frac{x_0^2}{k} + \frac{2 x_0}{k^2} + \frac{2}{k^3} \right)
    \]
    
    This method provides extreme precision, high numerical stability, and very fast execution speed for $x_0 > 0$.

    \item {Geometric and dimensional discretization:}\\
    Creation of a continuous search space $L \in [1, 100]\text{ nm}$ discretized over $N = 200$ uniform points for plotting curves, coupled with evaluations at specific discrete points for explicit summary tables.
\end{itemize}

\section{Numerical Results}

The calculations were performed with Python 3.8.1 using the NumPy library for numerical computations and Matplotlib for graphical visualization. The parameters for PbS are: \(E_g^{\text{bulk}}=0.41\,\mathrm{eV}\) \cite{Moreels2009}, \(m_e=0.08\,m_0\), \(m_h=0.1\,m_0\) where \(m_0 = 9.109\times10^{-31}\,\mathrm{kg}\) is the electron mass, \(n=10^{24}\,\mathrm{m}^{-3}\), \(T_c=300\,\mathrm{K}\), \(f_\omega=2.18\times10^{-5}\), \(P_{\text{inc}}=100\,\mathrm{mW/cm^2}\). Interactions are included with \(\theta=1\) (\(\Delta\epsilon_{\text{int}}=-0.22\,\mathrm{eV}\) \cite{Perdew1981}).

\subsection{Cube (volume \(L^3\))}

Table~\ref{tab:cube} gives the results for a cube of side \(L\). The maximum efficiency reaches 48.7\,\% at \(L = 5\ \mathrm{nm}\).

\begin{table}[H]
\centering
\caption{Cube PbS (volume \(L^3\)). With interactions (\(\theta=1\)). }
\label{tab:cube}
\begin{ruledtabular}
\hspace*{-2.5cm}\begin{tabular}{cccccccc}
\(L\) (nm) & \(E_g^{\text{eff}}\) (eV) & \(J_{sc}\) (mA/cm\(^2\)) & \(V_{oc}\) (V) & \(FF\) & \(\eta\) (\%) \\
\hline
3.0 & 3.23 & 4.9 &  2.87 & 0.95 & 13.5 \\
5.0 & 1.43 & 46.9 &  1.16 & 0.90 & 48.7 \\
7.0 & 0.93 & 73.9 & 0.70 & 0.85 & 43.6 \\
10.0 & 0.66 & 89.6 &  0.46 & 0.79 & 32.2 \\
14.0 & 0.54 & 96.6 &  0.34 & 0.74 & 24.7 \\
20.0 & 0.47 & 100.1 &  0.28 & 0.71 & 20.2 \\
50.0 & 0.42 & 102.8 &  0.24 & 0.68 & 16.5 \\
70.0 & 0.42 & 103.0 &  0.23 & 0.67 & 16.2 \\
100.0 & 0.41 & 103.2 &  0.23 & 0.67 & 16.0 \\
\end{tabular}
\end{ruledtabular}
\end{table}

\subsection{Square parallelepiped (fixed volume \(343\ \mathrm{nm}^3\))}

Table~\ref{tab:parallelo} presents the results for a square-base parallelepiped (\(a \times a \times b\)) of constant volume \(V_0 = 343\ \mathrm{nm}^3\) (cube of 7 nm). The efficiency optimum (49.0\,\%) is reached for \(b = 3.5\ \mathrm{nm}\).

\begin{table}[H]
\centering
\caption{Square parallelepiped (\(a \times a \times b\)), volume \(343\ \mathrm{nm}^3\). Results with interactions (\(\theta=1\)).}
\label{tab:parallelo}
\begin{ruledtabular}
\begin{tabular}{ccccccc}
\(b\) (nm) & \(a\) (nm) & \(E_g^{\text{eff}}\) (eV) & \(J_{sc}\) (mA/cm\(^2\)) & \(V_{oc}\) (V) & \(FF\) & \(\eta\) (\%) \\
\hline
3.5 & 9.90 & 1.27 & 54.5 & 1.02 & 0.88 & 49.0 \\
5.0 & 8.28 & 1.00 & 69.9 & 0.76 & 0.86 & 45.4 \\
7.0 & 7.00 & 0.93 & 73.9 & 0.70 & 0.85 & 43.6 \\
10.0 & 5.86 & 0.99 & 70.3 & 0.75 & 0.85 & 45.2 \\
14.0 & 4.95 & 1.14 & 61.4 & 0.90 & 0.87 & 48.1 \\
28.0 & 3.50 & 1.80 & 31.2 & 1.52 & 0.91 & 43.3 \\
50.0 & 2.62 & 2.88 & 8.0 & 2.53 & 0.94 & 19.2 \\
70.0 & 2.21 & 3.87 & 2.0 & 3.47 & 0.96 & 6.5 \\
100.0 & 1.85 & 5.34 & 0.2 & 4.87 & 0.97 & 0.9 \\
\end{tabular}
\end{ruledtabular}
\end{table}

\subsection{Cylinder (fixed volume \(343\ \mathrm{nm}^3\)) – with exact Bessel formula}

Table~\ref{tab:cylindre} gives the results for a cylinder of radius \(R\) and height \(H\) (same volume) using the exact confinement energy (Eq.~\eqref{eq:eg_cyl}). The efficiency optimum (49.0\,\%) is reached for \(H = 3.5\ \mathrm{nm}\).

\begin{table}[H]
\centering
\caption{Cylinder, volume \(343\ \mathrm{nm}^3\). Results with interactions and exact Bessel formula.}
\label{tab:cylindre}
\begin{ruledtabular}
\begin{tabular}{ccccccc}
\(H\) (nm) & \(R\) (nm) & \(E_g^{\text{eff}}\) (eV) & \(J_{sc}\) (mA/cm\(^2\)) & \(V_{oc}\) (V) & \(FF\) & \(\eta\) (\%) \\
\hline
3.5 & 5.59 & 1.26 & 55.2 & 1.01 & 0.88 & 49.0 \\
5.0 & 4.67 & 0.98 & 71.1 & 0.74 & 0.85 & 44.9 \\
7.0 & 3.95 & 0.90 & 75.5 & 0.67 & 0.84 & 42.7 \\
10.0 & 3.30 & 0.95 & 72.6 & 0.72 & 0.85 & 44.2 \\
14.0 & 2.79 & 1.09 & 64.5 & 0.85 & 0.87 & 47.3 \\
28.0 & 1.97 & 1.69 & 35.3 & 1.41 & 0.91 & 45.4 \\
50.0 & 1.48 & 2.68 & 10.5 & 2.35 & 0.94 & 23.1 \\
70.0 & 1.25 & 3.59 & 3.0 & 3.21 & 0.95 & 9.0 \\
100.0 & 1.04 & 4.95 & 0.4 & 4.50 & 0.96 & 1.6 \\
\end{tabular}
\end{ruledtabular}
\end{table}

\subsection{Sphere (radius \(R\))}

Table~\ref{tab:sphere} gives the results for a sphere of radius \(R\). The maximum efficiency reaches 49.1\,\% at \(R = 3\ \mathrm{nm}\). For large radii, the efficiency tends to the classical value.

\begin{table}[H]
\centering
\caption{Sphere (radius \(R\)), variable volume. Results with interactions (\(\theta=1\)).}
\label{tab:sphere}
\begin{ruledtabular}
\begin{tabular}{ccccccc}
\(R\) (nm) & \(V\) (nm\(^3\)) & \(E_g^{\text{eff}}\) (eV) & \(J_{sc}\) (mA/cm\(^2\)) & \(V_{oc}\) (V) & \(FF\) & \(\eta\) (\%) \\
\hline
3.0 & 113.1 & 1.35 & 50.5 & 1.09 & 0.89 & 49.1 \\
4.0 & 268.1 & 0.94 & 73.2 & 0.71 & 0.85 & 43.9 \\
4.3 & 333.0 & 0.87 & 77.5 & 0.64 & 0.84 & 41.5 \\
5.0 & 523.6 & 0.75 & 84.6 & 0.53 & 0.81 & 36.5 \\
6.0 & 904.8 & 0.65 & 90.7 & 0.44 & 0.78 & 31.1 \\
7.0 & 1436.8 & 0.58 & 94.2 & 0.38 & 0.76 & 27.4 \\
10.0 & 4188.8 & 0.49 & 99.0 & 0.30 & 0.72 & 21.7 \\
20.0 & 33510.3 & 0.43 & 102.3 & 0.25 & 0.68 & 17.3 \\
50.0 & 523598.8 & 0.41 & 103.2 & 0.23 & 0.67 & 15.9 \\
70.0 & 1436755.0 & 0.41 & 103.2 & 0.23 & 0.67 & 15.9 \\
100.0 & 4188790.2 & 0.41 & 103.2 & 0.23 & 0.67 & 15.9 \\
\end{tabular}
\end{ruledtabular}
\end{table}

\subsection{Comparative graphs}

\begin{figure}[H]
\centering
\hspace*{-0.1cm}\includegraphics[width=0.5\textwidth]{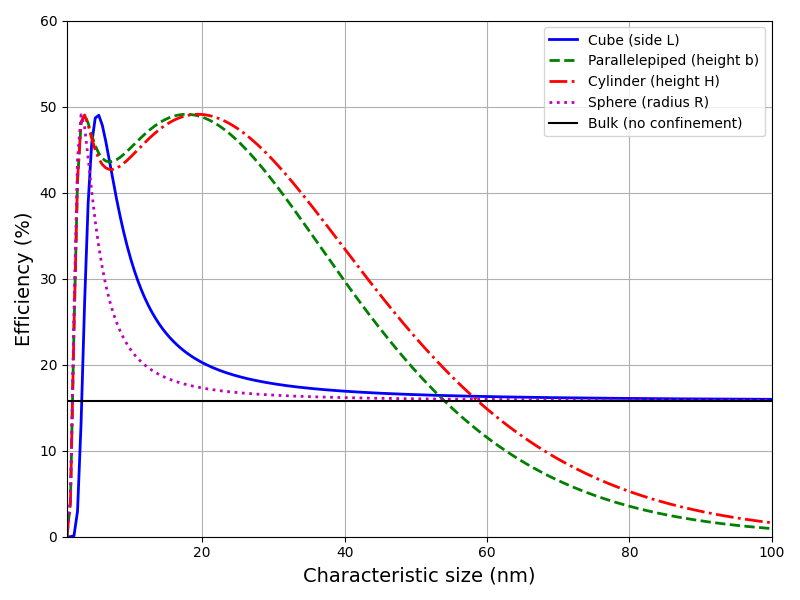}
\caption{Simulation of the efficiency of PbS nanostructured solar cells as a function of the characteristic size. The curves for the parallelepiped and cylinder are plotted at constant volume \(V_0 = 343\ \mathrm{nm}^3\). The horizontal line corresponds to the efficiency of bulk PbS (15.8\,\%). Silicon records are around 27\,\%.}
\label{fig:fig1}
\end{figure}

Figure~\ref{fig:fig1} shows the evolution of efficiency as a function of the characteristic size for the four geometries (cube: side \(L\); square parallelepiped: height \(b\); cylinder: height \(H\); sphere: radius \(R\)). A sharp peak is observed around 5 nm for the cube, while the constant-volume shapes (parallelepiped and cylinder) exhibit two peaks: one for very small heights (<5 nm) and a second around 14 nm. The sphere has an optimum at \(R=3\) nm (49.1\,\%) and then decreases rapidly. The horizontal dashed line corresponds to the efficiency of bulk PbS (15.8\,\%). Silicon records are around 27\,\%.

\begin{figure}[H]
\centering
\hspace*{-0.1cm}\includegraphics[width=0.5\textwidth]{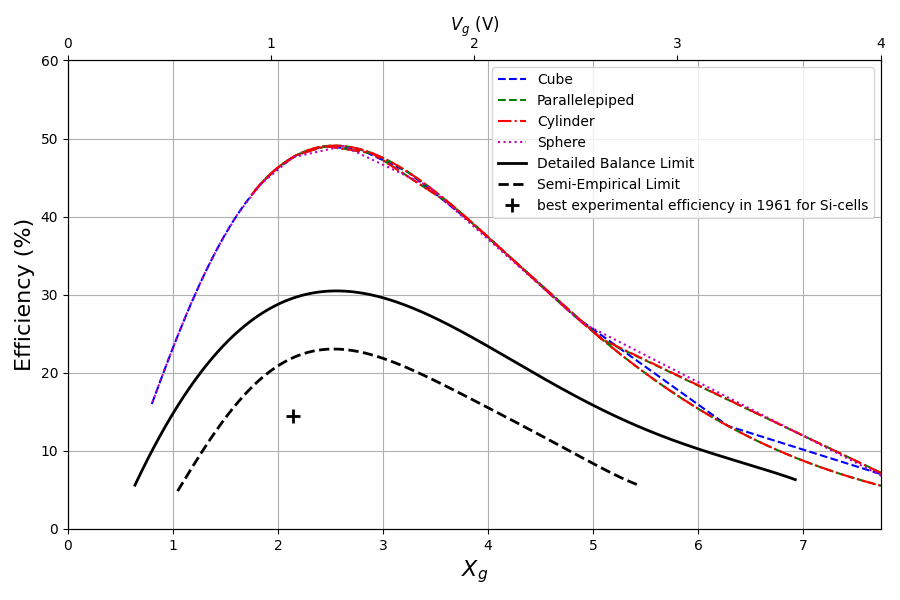}
\caption{Comparison of the ``semi-empirical limit'' of solar cell efficiency with the ``detailed balance limit'' \cite{Shockley1961}, and simulations for PbS nanostructures.}
\label{fig:fig2}
\end{figure}

Figure~\ref{fig:fig2} compares the efficiencies of the nanostructures with the classical theoretical limits (semi-empirical limit and detailed balance limit).

In the original Shockley-Queisser limit \cite{Shockley1961}, the dimensionless parameter
\[
x_g = \frac{qV_g}{k_B T_s} = \frac{E_g}{k_B T_s},
\]
is defined, where \(V_g = E_g/q\) is the voltage corresponding to the bandgap, \(q\) the elementary charge, \(k_B\) the Boltzmann constant, and \(T_s = 6000\,\mathrm{K}\) the temperature of the sun. This parameter appears in the calculation of the ultimate efficiency and the saturation current.

\begin{figure}[H]
\centering
\includegraphics[width=0.5\textwidth]{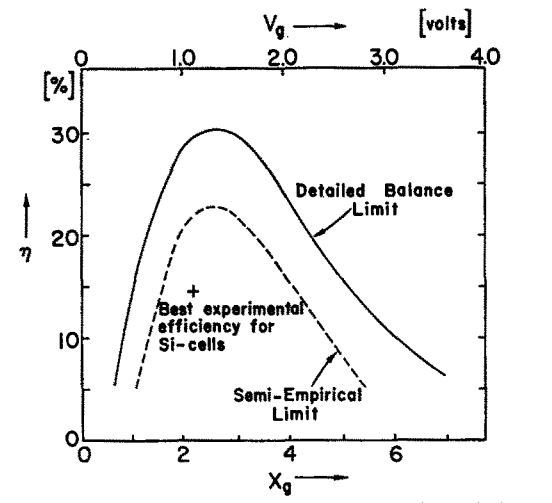}
\caption{Comparison of the ``semi-empirical limit'' of solar cell efficiency with the ``detailed balance limit'' derived in that article. The '+' symbols indicate the ``best experimental efficiency of the time'' for silicon cells.}
\label{fig:fig3}
\end{figure}

Figure~\ref{fig:fig3} is an additional comparison taken from the original reference \cite{Shockley1961}.

\section{Discussion}

The results extended up to 100 nm confirm the trends observed for small sizes and reveal interesting asymptotic behaviours. For the cube, the maximum efficiency of 48.7\,\% at 5 nm exceeds the classical Shockley-Queisser limit by more than 15 percentage points. This gain comes from the widening of the bandgap (1.43 eV) and the modification of the saturation current via \(\Delta\) \cite{Ranaivoson2022,Ravelonjato2023}. For very small sizes, \(L = 3\) nm, the gap becomes too large (3.23 eV), which strongly reduces solar absorption and gives an efficiency of 13.5\,\%. At the other end, \(L = 100\) nm, the efficiency drops to 16\,\%, close to that of a bulk PbS cell.

For constant-volume shapes, two optima appear: one for very flattened shapes, \(b = 3.5\) nm, \(H = 3.5\) nm with 49.0\,\%, and a second for moderately elongated shapes, \(b = 14\) nm, \(H = 14\) nm with 48.1\,\% and 47.3\,\% respectively. These two peaks correspond to compromises between \(J_{sc}\) and \(V_{oc}\). Unlike the cube and sphere, the curves for the parallelepiped and cylinder are plotted at constant volume. When the height \(b\) of the parallelepiped or \(H\) of the cylinder becomes very large, the transverse section \(a\) or \(R\) decreases as \(1/\sqrt{b}\) or \(1/\sqrt{H}\) to keep the volume fixed. The confinement in the lateral directions then becomes extremely strong, which considerably increases the confinement energy and thus the effective bandgap \(E_g^{\text{eff}}\). As a consequence, the short-circuit current \(J_{sc}\) drops drastically and the efficiency tends to zero. Conversely, for very small heights, the axial confinement dominates, and an efficiency optimum appears. This bell-shaped behaviour with a zero asymptote on both sides is characteristic of fixed-volume systems. The sphere exhibits a very sharp optimum at \(R = 3\) nm with an efficiency of 49.1\,\%, the highest among all the studied geometries. For larger radii, the efficiency drops rapidly because the gap decreases and the voltage collapses. The sphere thus seems particularly promising for very small quantum dots (3 nm). For comparison, bulk PbS reaches only about 15.8\,\% efficiency, highlighting the importance of quantum confinement. This behaviour is consistent with the predictions of the same QPS formalism for anisotropic pressure in confined Fermi gases \cite{Ravelonjato2026}.

During the continuous variation of the size \( L \), a clear splitting of the efficiency curves for the cylinder and the parallelepiped is observed, particularly visible in the strong confinement region (\( X_g \geq 4.5 \)). This phenomenon is explained by the non-monotonic nature of quantum confinement under the constraint of a constant volume (\( V_0 = 343 \, \text{nm}^3 \)). Since the confinement equation presents an absolute minimum (\( X_g \approx 1.06 \) for the cylinder at \( H \approx 7.18 \, \text{nm} \), and \( X_g \approx 1.08 \) for the parallelepiped at \( b = 7 \, \text{nm} \)), any bandgap value above this threshold admits two distinct geometric solutions.

Comparing our results with the classical theoretical limits (Fig.~\ref{fig:fig2}), all optimised nanostructures largely exceed the detailed balance limit of Shockley-Queisser, about 40\,\% at best for a gap of 1.4 eV, and even more the semi-empirical limit, about 30\,\%. This confirms that quantum confinement in low-dimensional materials offers a promising route to surpass the traditional limits of photovoltaics. These results are consistent with Anttu's predictions for nanowires, where a reduction of photon emission into the substrate allows the open-circuit voltage to be increased beyond the bulk limit \cite{Anttu2015}. Moreover, the introduction of the external radiative efficiency (ERE) in modern analyses \cite{Seo2025} allows the quantification of non-radiative losses, which in our model are partially captured by the coefficient \(\theta\) and the exchange-correlation correction. As emphasised in Markvart's review \cite{Markvart2022}, these extensions of the SQ model are essential for understanding and exceeding the limits of conventional devices.

The results of our model (about 49\,\% for a 3 nm sphere) are comparable to the predictions of the detailed balance limit for tandem structures: De Vos showed that a stack of two cells can reach 42\,\%, three cells 49\,\% and an infinite number of cells 68\,\% under unconcentrated illumination \cite{DeVos1980}. This underlines that quantum confinement offers performances equivalent to complex multi-junction architectures, but with a single nanostructured junction.

The best commercial silicon solar cells reach efficiencies of 27.09\,\% \cite{Wang2024} and 27.0\,\% \cite{Xie2025}, well below the 49\,\% predicted here for optimised nanostructures, cube of 5 nm and sphere of 3 nm. This contrast illustrates the considerable potential of quantum confinement to surpass the limits of conventional devices, even if the industrial fabrication of such nanostructures remains a challenge.

Our model assumes an ideal junction and a single-size nanostructure. In practice, size distribution, defects, and series resistances will reduce the observed efficiencies \cite{Sahin2018,Alexandre2025}. Nevertheless, the obtained values constitute realistic upper bounds for PbS quantum dot solar cells \cite{Moreels2009,Etgar2013}. The existence of two peaks for constant-volume shapes and a very sharp peak for the sphere suggests that several geometric configurations can achieve efficiencies above 48\,\%, which is very encouraging for device design.

\section{Conclusion}

We have developed an extension of the Shockley-Queisser limit for nanostructured solar cells using the quantum phase space formalism. The unified parameter \(B_{ll}\) relates the size and shape of the nanostructures to the effective density of states and the saturation current. For the cylindrical geometry, we introduced the exact confinement energy based on the first zero of the Bessel function. For the sphere, we used the standard confinement energy. Numerical simulations for PbS, extended up to 100 nm, show that efficiencies above 48\,\% are possible for 5 nm cubes, flattened constant-volume shapes, and especially for 3 nm spheres. A second efficiency peak appears for heights around 14 nm. Beyond 30 nm, the efficiency drops rapidly to values below 20\,\%. For comparison, the same material without confinement would give only about 15.8\,\%. The considerable gap, up to 33 percentage points, illustrates the interest of quantum confinement for photovoltaics. These results far exceed the classical Shockley-Queisser limits and open the way to the geometric optimisation of quantum dot-based solar cells. Electron-electron interactions have a moderate but non-negligible effect. This work provides a theoretical framework for the design of very high-efficiency devices.

\section*{Data Availability Statement}
The data that support the findings of this study are available within the article and its supplementary material.

\begin{acknowledgments}
The authors acknowledge the support of INSTN and CEA-Liten. 
\end{acknowledgments}

\appendix

\section{Determination of \(B_{ll}\) by limit identification}
\label{app:A}

\subsection{Discrete grand potential}
\begin{equation}
\Omega_{\text{disc}} = -\frac{g_s k_B T}{8} \sum_{j=1}^{\infty} \frac{(-1)^{j+1} \xi^j}{j} \frac{1}{\prod_{l=1}^3 \sinh\left(j \frac{\beta B_{ll}}{m^*}\right)},
\label{eq:grand_pot}
\end{equation}
where \(\xi = e^{\beta\mu}\) is the fugacity, \(\beta = 1/(k_B T)\), and \(\mu\) the chemical potential.

\subsection{Classical limit – general case}
In the classical limit (\(\xi \ll 1\), \(\beta B_{ll}/m^* \ll 1\)), only \(j=1\) contributes and \(\sinh x \approx x\). We obtain:
\begin{equation}
\Omega_{\text{disc}} \approx -\frac{g_s k_B T}{8} \xi \frac{(m^*)^3}{\beta^3 B_{xx} B_{yy} B_{zz}},
\label{eq:Omega_disc_class}
\end{equation}
and the continuous counterpart:
\begin{equation}
\Omega_{\text{cont}} \approx -g_s k_B T \frac{V}{\lambda_{\text{th}}^3} \xi.
\label{eq:Omega_cont_class}
\end{equation}
Identification \(\Omega_{\text{disc}} = \Omega_{\text{cont}}\) gives:
\begin{equation}
B_{xx} B_{yy} B_{zz} = \frac{(m^*)^3 \lambda_{\text{th}}^3}{8\beta^3 V}.
\label{eq:prodB_class}
\end{equation}

\subsection{Expressions for specific geometries}
\begin{itemize}
\item \textbf{Cube} (\(V = L^3\), \(B_{xx}=B_{yy}=B_{zz}=B\)):
  \[
  B^3 = \frac{(m^*)^3 \lambda_{\text{th}}^3}{8\beta^3 L^3} \;\Rightarrow\; B = \frac{m^*}{2\beta}\frac{\lambda_{\text{th}}}{L} = \frac{\hbar}{2L}\sqrt{2\pi m^* k_B T}.
  \]
  We identify \(\Delta = \sqrt{2\pi m^* k_B T}\) (classical limit).
\item \textbf{Sphere} (\(V = \frac{4}{3}\pi R^3\), isotropy \(B_{xx}=B_{yy}=B_{zz}=B\)):
  \[
  B^3 = \frac{(m^*)^3 \lambda_{\text{th}}^3}{8\beta^3 (4\pi/3)R^3} \;\Rightarrow\; B = \frac{\hbar}{2(4\pi/3)^{1/3} R}\sqrt{2\pi m^* k_B T}.
  \]
  Setting \(L_{\text{sph}} = (4\pi/3)^{1/3}R\), we have \(B = \frac{\hbar}{2L_{\text{sph}}}\Delta\).
\item \textbf{Cylinder} (with \(L_x = L_y = \sqrt{\pi}R\), \(L_z = H\)):
  \[
  B_{xx}B_{yy}B_{zz} = \left(\frac{\hbar}{2\sqrt{\pi}R}\Delta\right)^2 \left(\frac{\hbar}{2H}\Delta\right) = \frac{\hbar^3}{8\pi R^2 H}\Delta^3.
  \]
  Since \(V = \pi R^2 H\), \(\prod B_{ll} = \frac{\hbar^3}{8V}\Delta^3\). Identifying with Eq.~\eqref{eq:prodB_class} where \(\Delta = \sqrt{2\pi m^* k_B T}\) in the classical limit verifies consistency.
\end{itemize}

\subsection{Quantum degenerate limit (\(T\to 0\))}
From the discrete particle density:
\begin{equation}
n_{\text{disc}} \approx \frac{g_s}{V} \frac{1}{8} \left(\frac{m^*}{2B}\right)^3 \mu^3,
\label{eq:ndisc}
\end{equation}
and the continuous density \(n = \frac{g_s}{6\pi^2}\left(\frac{2m^*\epsilon_F}{\hbar^2}\right)^{3/2}\), we obtain for a cube:
\begin{equation}
B = \frac{\hbar}{2L}\sqrt{\pi\hbar^2\left(6\pi^2\frac{n}{g_s}\right)^{2/3}}.
\label{eq:B_quantum}
\end{equation}
This shows that \(\Delta\) must be replaced by \(\sqrt{\pi\hbar^2(6\pi^2 n/g_s)^{2/3}}\) in this limit. The unified expression (Eq.~\eqref{eq:delta}) interpolates between the two regimes.

\section{Correction of the saturation current by the product of \(B_{ll}\)}
\label{app:B}
\begin{equation}
\zeta_{\text{QPS}} = \prod_{l=1}^{3} \frac{1}{2\sinh(\beta B_{ll}/m)} \approx \frac{m^3}{8\beta^3 \prod B_{ll}},
\label{eq:zeta}
\end{equation}
\begin{equation}
\frac{J_0^{\text{QD}}}{J_0^{\text{bulk}}(E_g^{\text{eff}})} = \frac{\zeta_{\text{QPS}}}{\zeta_{\text{class}}} = \frac{\lambda_{\text{th}}^3 m^3}{8\beta^3 V \prod B_{ll}}.
\label{eq:ratioJ0}
\end{equation}

\section{Numerical calculation details}
\label{app:C}

The numerical simulations were carried out using Python 3.8.1. The NumPy library was employed for all array operations and numerical processing, while the Matplotlib package was used to generate the figures presented in this work.

The core integral appearing in Eqs.~\eqref{eq:j0bulk} and \eqref{eq:jsc}, defined as
\begin{equation}
I(x_0) = \int_{x_0}^{\infty} \frac{x^2}{e^x - 1} \, dx,
\label{eq:I_def}
\end{equation}
was not evaluated using adaptive quadrature methods. Instead, it was computed using the exact convergent series expansion derived from the identity \(1/(e^x - 1) = \sum_{k=1}^{\infty} e^{-kx}\):
\begin{equation}
I(x_0) = \sum_{k=1}^{\infty} e^{-k x_0} \left( \frac{x_0^2}{k} + \frac{2 x_0}{k^2} + \frac{2}{k^3} \right).
\label{eq:I_series}
\end{equation}

For \(x_0 > 0\), this series converges exponentially. The summation was performed up to \(k_{\text{max}} = 100\), which ensures a numerical precision better than \(10^{-15}\) for all relevant values of \(x_0\) encountered in this study. For the special case \(x_0 = 0\), the integral reduces to \(I(0) = 2\zeta(3) \approx 2.404113806\), where \(\zeta\) denotes the Riemann zeta function.

The confinement energies for the different geometries (cube, parallelepiped, cylinder, and sphere) were implemented exactly as defined in Section~\ref{sec:qps}, using the first zero of the Bessel function \(j_{0,1} = 2.4048255577\) for the cylindrical case. The search space for the characteristic size \(L\) was discretized over the interval \([1, 100]\) nm with \(N = 200\) uniformly spaced points for generating the continuous curves (Figs.~\ref{fig:fig1}--\ref{fig:fig3}). Supplementary evaluations were performed at selected discrete sizes to construct the summary tables (Tables~\ref{tab:cube}--\ref{tab:sphere}). All physical constants used in the simulation are those listed in Section~\ref{sec:qps}.

\section{Verification of the return to classical values for \(L \to \infty\)}
\label{app:D}

\begin{enumerate}
\item \(E_g^{\text{eff}} \to E_g^{\text{bulk}}\) since \(\Delta E_{\text{conf}} \propto 1/L^2\).
\item \(B \sim \frac{\hbar}{2L}\sqrt{2\pi m k_B T_c}\).
\item \(\frac{\lambda_{\text{th}} m}{2\beta L B} \to 1\) \(\Rightarrow\) \(J_0^{\text{QD}} \to J_0^{\text{bulk}}\).
\item \(J_{sc} \to q A_p t_s f_\omega Q_s(E_g^{\text{bulk}})\), \(V_{oc}\), \(FF\) and \(\eta\) tend to the classical values.
\end{enumerate}

\section{Parametrisation of the correlation energy (Perdew-Zunger)}
\label{app:E}

\begin{equation}
\left.
\begin{aligned}
\epsilon_c(r_s,\xi) &= \epsilon_c^U(r_s) + f(\xi)\bigl[\epsilon_c^P(r_s) - \epsilon_c^U(r_s)\bigr], \\
f(\xi) &= \frac{(1+\xi)^{4/3}+(1-\xi)^{4/3}-2}{2^{4/3}-2}
\end{aligned}
\right\}
\label{eq:energie}  
\end{equation}
where \(r_s = (3/(4\pi n))^{1/3}\) is the density parameter, \(\xi = (n_\uparrow - n_\downarrow)/n\) the spin polarisation, \(U\) denotes the unpolarised case (\(\xi=0\)) and \(P\) the fully polarised case (\(\xi=1\)).

For \(r_s \ge 1\):
\begin{equation}
\epsilon_c^i(r_s) = \frac{\gamma_i}{1 + \beta_1^i \sqrt{r_s} + \beta_2^i r_s}.
\end{equation}
For \(r_s < 1\):
\begin{equation}
\epsilon_c^i(r_s) = A_i \ln r_s + B_i + C_i r_s \ln r_s + D_i r_s.
\end{equation}

Constants (Ceperley-Alder):
\begin{center}
\begin{ruledtabular}
\begin{tabular}{c|cc}
Parameter & \(U\) & \(P\) \\
\hline
\(\gamma\) & -0.1423 & -0.0843 \\
\(\beta_1\) & 1.0529 & 1.3981 \\
\(\beta_2\) & 0.3334 & 0.2611 \\
\(A\) & 0.0311 & 0.01555 \\
\(B\) & -0.048 & -0.0269 \\
\(C\) & 0.0020 & 0.0007 \\
\(D\) & -0.0116 & -0.0048 \\
\end{tabular}
\end{ruledtabular}
\end{center}

Exchange energy:
\begin{equation}
\epsilon_x(n) = -\frac{3e^2}{4\pi}\left(\frac{3n}{\pi}\right)^{1/3}.
\end{equation}

\section{Cube PbS (volume \(L^3\)). With interactions (\(\theta=1\)). Columns \(J_0^{\text{no}}\) and \(J_0^{\text{w}}\) correspond to \(\theta=0\) and \(\theta=1\), respectively.}
\label{app:F}

\bgroup  
\renewcommand{\thetable}{F}  

\begin{table}[H]
\centering  
\caption{Cube PbS (volume \(L^3\)). With interactions (\(\theta=1\)). Columns \(J_0^{\text{no}}\) and \(J_0^{\text{w}}\) correspond to \(\theta=0\) and \(\theta=1\), respectively.}
\label{tab:cube}

\makebox[\linewidth][l]{
  \hspace*{-0.5cm}
  \begin{ruledtabular}
  \begin{tabular}{cccccccc}
  \(L\) (nm) & \(E_g^{\text{eff}}\) (eV) &  \(J_0^{\text{no}}\) (A/cm\(^2\)) & \(J_0^{\text{w}}\) (A/cm\(^2\)) & \(V_{oc}\) (V) & \(FF\) & \(\eta\) (\%) \\
  \hline
  3.0 & 3.23 & 1.14e-51 & 3.49e-51 & 2.87 & 0.95 & 13.5 \\
  5.0 & 1.43 & 4.75e-22 & 1.45e-21 & 1.16 & 0.90 & 48.7 \\
  7.0 & 0.93 & 4.63e-14 & 1.42e-13 & 0.70 & 0.85 & 43.6 \\
  10.0 & 0.66 & 6.65e-10 & 2.03e-09 & 0.46 & 0.79 & 32.2 \\
  14.0 & 0.54 & 5.48e-08 & 1.68e-07 & 0.34 & 0.74 & 24.7 \\
  20.0 & 0.47 & 5.50e-07 & 1.68e-06 & 0.28 & 0.71 & 20.2 \\
  50.0 & 0.42 & 3.45e-06 & 1.06e-05 & 0.24 & 0.68 & 16.5 \\
  70.0 & 0.42 & 4.09e-06 & 1.25e-05 & 0.23 & 0.67 & 16.2 \\
  100.0 & 0.41 & 4.48e-06 & 1.37e-05 & 0.23 & 0.67 & 16.0 \\
  \end{tabular}
  \end{ruledtabular}
}%

\end{table}
\egroup  
\bibliography{mybib}  


\end{document}